\documentclass[aps,prc,reprint,groupedaddress, nofootinbib]{revtex4-2}

\usepackage{graphicx} 
\usepackage{amsmath}
\usepackage{amssymb}
\usepackage{ifthen}
\usepackage{xspace}
\NewDocumentCommand{\nuc} {m m o} {\ensuremath{{^{#2}\IfNoValueTF{#3}{}{_{#3}}\textrm{#1}}}\xspace}
\NewDocumentCommand{\electron} {} {\ensuremath{e^-}\xspace}
\usepackage{lineno}
\usepackage{hyperref}
\usepackage{float}
\usepackage{upgreek}
\usepackage{acro}

\DeclareAcronym{omc}{short = OMC, long = ordinary muon capture}
\DeclareAcronym{bb}{short = \ensuremath{\beta\beta}, long = double beta}
\DeclareAcronym{onbb}{short = \ensuremath{0\nu\beta\beta}, long = neutrinoless double beta}

\DeclareAcronym{nme}{short = NME, long = nuclear matrix element}
\DeclareAcronym{qrpa}{short = QRPA, long = quasiparticle random phase approximation}

\DeclareAcronym{monument}{short = MONUMENT, long = Muon Ordinary capture for NUclear Matrix elemENTs calculations}
\DeclareAcronym{psi}{short = PSI, long = Paul Scherrer Institute}

\DeclareAcronym{pmt}{short = PMT, long = photo multiplier tube}
\DeclareAcronym{hpge}{short = HPGe, long = high-purity germanium}
\DeclareAcronym{bege}{short = BEGe, long = broad energy germanium}
\DeclareAcronym{coax}{short = COAX, long = (semi-)coaxial}
\DeclareAcronym{fep}{short = FEP, long = full-energy peak}

\DeclareAcronym{daq}{short = DAQ, long = data acquisition}
\DeclareAcronym{midas}{short = MIDAS, long = Maximum Integration Data Acquisition System}
\DeclareAcronym{alpaca}{short = ALPACA, long = Applied Llama Program for ACcelerator Applications}
\DeclareAcronym{lf}{short = LF, long = low frequency}
\DeclareAcronym{hf}{short = HF, long = high frequency}

\DeclareAcronym{mc}{short = MC, long = Monte Carlo}

\begin{document}


\title{Determination of the Muon Lifetime in \(^{76}\)Se with the MONUMENT experiment}



\author{G.~R.~Araujo}
\affiliation{Physik-Institut, University of Zurich, Zurich, Switzerland}

\author{D.~Bajpai}
\affiliation{Department of Physics and Astronomy, University of Alabama, Tuscaloosa, AL, USA}

\author{L.~Baudis}
\affiliation{Physik-Institut, University of Zurich, Zurich, Switzerland}

\author{V.~Belov}
\affiliation{Joint Institute for Nuclear Research, Dubna, Russia}

\author{E.~Bossio}
\altaffiliation{currently at IRFU, CEA, Université Paris-Saclay, Gif-sur-Yvette, France}
\affiliation{Technical University of Munich, Garching, Germany}

\author{T.E.~Cocolios}
\affiliation{KU Leuven, Institute for Nuclear and Radiation Physics, Leuven, Belgium}

\author{H.~Ejiri}
\affiliation{Research Center on Nuclear Physics, Osaka University, Ibaraki, Osaka, Japan}

\author{M.~Fomina}
\affiliation{Joint Institute for Nuclear Research, Dubna, Russia}

\author{K.~Gusev}
\affiliation{Joint Institute for Nuclear Research, Dubna, Russia}
\affiliation{Technical University of Munich, Garching, Germany}

\author{I.H.~Hashim}
\affiliation{Department of Physics, Faculty of Science, Universiti Teknologi Malaysia, Johor Bahru, Johor, Malaysia}

\author{M.~Heines}
\affiliation{KU Leuven, Institute for Nuclear and Radiation Physics, Leuven, Belgium}

\author{S.~Kazartsev}
\affiliation{Joint Institute for Nuclear Research, Dubna, Russia}

\author{A.~Knecht}
\affiliation{Paul Scherrer Institut, Villigen, Switzerland}

\author{E.~Mondrag\'on}
\email[]{elizabeth.mondragon@tum.de}
\altaffiliation{currently at Max-Planck-Institut für Kernphysik, Heidelberg, Germany}
\affiliation{Technical University of Munich, Garching, Germany}

\author{Z.W.~Ng}
\affiliation{Department of Physics, Faculty of Science, Universiti Teknologi Malaysia, Johor Bahru, Johor, Malaysia}

\author{I.~Ostrovskiy}
\affiliation{Institute of High Energy Physics, Chinese Academy of Sciences, Beijing 100049, China}

\author{N.~Rumyantseva}
\affiliation{Joint Institute for Nuclear Research, Dubna, Russia}
\affiliation{Technical University of Munich, Garching, Germany}

\author{S.~Sch\"{o}nert}
\affiliation{Technical University of Munich, Garching, Germany}

\author{M.~Schwarz}
\affiliation{Technical University of Munich, Garching, Germany}

\author{A.~Shehada}
\affiliation{Joint Institute for Nuclear Research, Dubna, Russia}

\author{E.~Shevchik}
\affiliation{Joint Institute for Nuclear Research, Dubna, Russia}

\author{M.~Shirchenko}
\affiliation{Joint Institute for Nuclear Research, Dubna, Russia}

\author{Y.~Shitov}
\affiliation{Institute of Experimental and Applied Physics, Czech Technical University in Prague, Prague, Czech Republic}

\author{J.~Suhonen}
\affiliation{Department of Physics, University of  Jyväskylä, Jyväskylä, Finland}
\affiliation{International Centre for Advanced Training and Research in Physics (CIFRA), Bucharest-Magurele, Romania\\}

\author{S.M.~Vogiatzi}
\affiliation{KU Leuven, Institute for Nuclear and Radiation Physics, Leuven, Belgium}

\author{C.~Vogl}
\affiliation{Technical University of Munich, Garching, Germany}

\author{C.~Wiesinger}
\altaffiliation{currently at Max-Planck-Institut für Kernphysik, Heidelberg, Germany}
\affiliation{Technical University of Munich, Garching, Germany}

\author{I.~Zhitnikov}
\email[]{igorzhitnikov@jinr.ru}
\affiliation{Joint Institute for Nuclear Research, Dubna, Russia}

\author{D.~Zinatulina}
\affiliation{Joint Institute for Nuclear Research, Dubna, Russia}

\collaboration{MONUMENT collaboration}
\noaffiliation

\date{\today}

\begin{abstract}
\Acl*{omc} provides a benchmark for the nuclear physics models of \acl*{onbb} decay under comparable momentum transfer conditions.
The total capture strength defines the lifetime of the muonic atom.
The muon lifetime in \nuc{Se}{76}, the daughter nucleus of \nuc{Ge}{76}, was determined with improved accuracy by the \acs*{monument} collaboration, using an array of \acl*{hpge} detectors and a set of scintillator counters at the \(\pi\)E1 muon beam line of the \acl*{psi}.
The new value of (\(135.1 \pm 0.5\))~ns agrees with phenomenological calculations based on the \acl*{qrpa} with unquenched axial-vector coupling.
\end{abstract}


\maketitle


\section{\label{sec:intro}Introduction}

\Ac{onbb} decay is a hypothesized nuclear process in which a nucleus \nuc{X}{A}[Z] undergoes \ac{bb} decay to a daughter nucleus \nuc{Y}{A}[Z+2], emitting two electrons and no antineutrinos:
\begin{equation}
 \nuc{X}{A}[Z] \rightarrow \nuc{Y}{A}[Z+2]+ 2 \electron.
\label{eq:onbb-decay}
\end{equation}
It violates lepton number conservation, which could be linked to the matter-antimatter asymmetry of our universe~\cite{Fukugita:1986hr} and would evidence the Majorana nature of neutrinos, implying that they are their own antiparticles~\cite{Schechter:1981bd}.
Its half-life, \(T_{1/2}\), depends on unknown particle physics properties, such as the effective Majorana neutrino mass, \(m_{\beta\beta}\), when the decay is mediated by light Majorana neutrino exchange; on the corresponding nuclear physics quantities, such as the \ac{nme}, \(\mathcal{M}\), and the effective axial-vector coupling constant, \(g_A\); and on the phase space factor, \(\mathcal{G}\):
\begin{equation}
  \frac{1}{T_{1/2}} = \mathcal{G} \cdot g_A^4 \left| \mathcal{M} \right|^2 \cdot \frac{m_{\beta\beta}^2}{m_e^2}.
\label{eq:onbb-rate}
\end{equation}
To this date, the \acp{nme} computed by different groups using different many-body methods exhibit a significant spread of factors 2 to 3, constituting the largest source of uncertainty when relating \ac{onbb} decay constraints to particle physics properties and when comparing experiments using different \ac{bb} isotopes~\cite{Gomez-Cadenas:2023vca, Agostini:2022zub}.

\begin{figure}
\includegraphics[width=0.65\columnwidth]{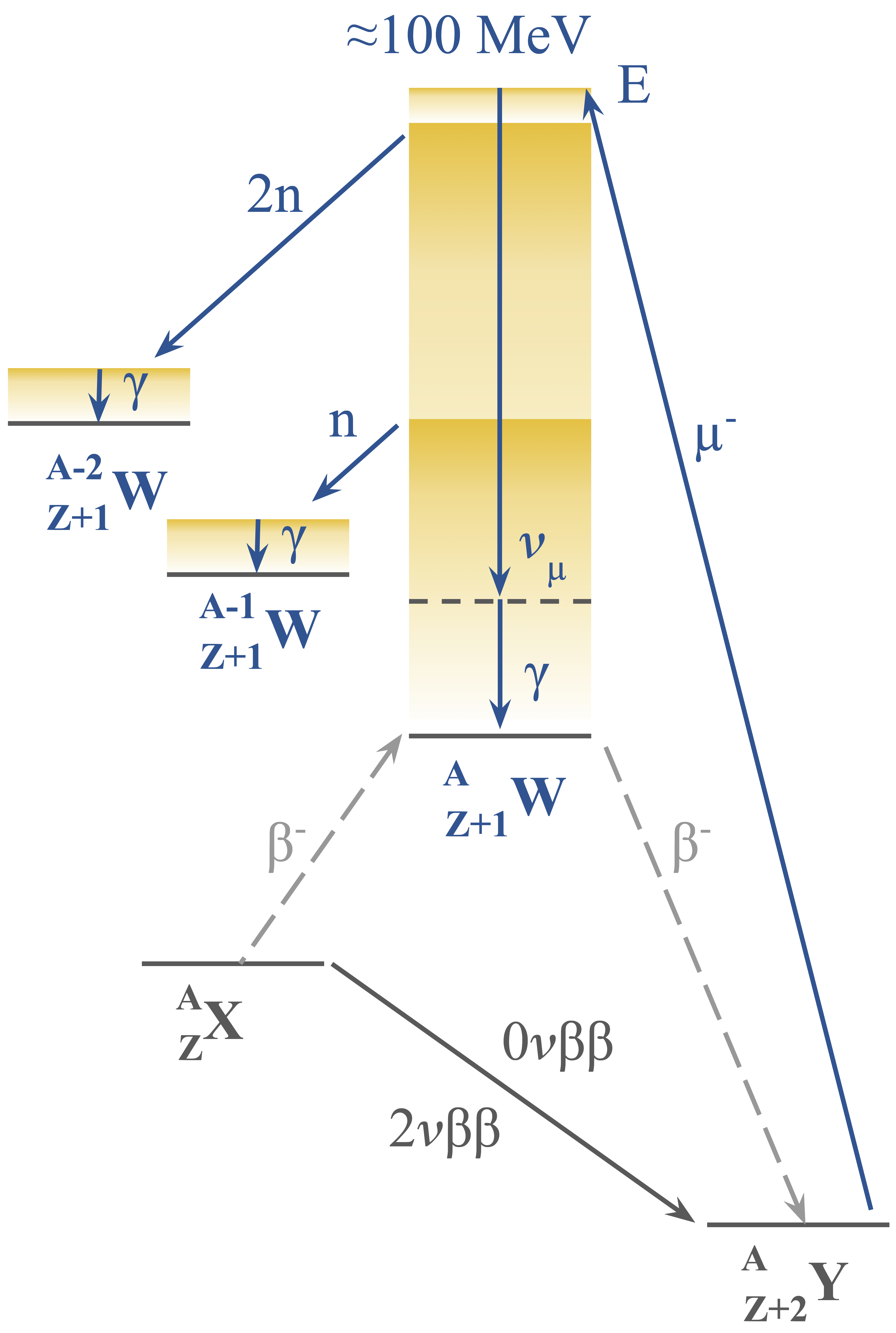}
\caption{\label{fig:omc-scheme}
Schematic comparison of \acs*{omc} and \acs*{bb} decay. 
\acs*{omc} on \nuc{Y}{A}[Z+2] provides experimental access to the right virtual transition of \acs*{bb} decay, populating the intermediate nucleus \nuc{W}{A}[Z+1] under similar momentum transfer as \acs*{onbb} decay.
The subsequent $\upgamma$-ray emission after \acs*{omc} allows to tag the capture process and determine the muon lifetime in the target nucleus.
For the \ac*{bb}-decay isotope \nuc{Ge}{76}, the daughter isotope and \acs*{omc} target is \nuc{Se}{76}, which transforms into \nuc{As}{76} after \acs*{omc} and other \nuc{As}{} isotopes after neutron emission.
}
\end{figure}

\Ac{omc} on \nuc{Y}{A}[Z+2] offers a benchmark for the nuclear physics models of \ac{onbb} decay under similar momentum transfer conditions~\cite{Kortelainen:2002bz}, populating the intermediate nucleus \nuc{W}{A}[Z+1]:
\begin{equation}
 \mu^- +  \nuc{Y}{A}[Z+2] \rightarrow \nuc{W}{A}[Z+1]^* + \nu_\mu
\label{eq:omc}
\end{equation}
The subsequent de-excitation of \(\nuc{W}{A}[Z+1]^*\), predominantly by the emission of neutrons, protons or gamma ($\upgamma$) rays, reveals information about the nucleus under a momentum transfer of about 100~MeV/c.
\Ac{omc} can be considered the experimentally accessible analogue of the right virtual transition of \ac{onbb} decay~\cite{Ejiri:2019ezh}.
Fig.~\ref{fig:omc-scheme} shows a schematic comparison of both processes.
The lifetime of the muonic atom until the muon decays or is captured, referred to as muon lifetime, provides information about the total capture strength.
It can be measured with high experimental accuracy and allows comparison with theoretical calculations performed within the same nuclear models used to compute the \acp{nme} of \ac{onbb} decay.

The \acs{monument} collaboration conducted a series of \ac{omc} measurements at the \ac{psi} in Switzerland, using mainly targets that are relevant for \ac{onbb} decay, such as \nuc{Ti}{48}, \nuc{Se}{76} and \nuc{Ba}{136}~\cite{Araujo:2024mdu}.
Earlier works of a similar type have been presented in \cite{Zinatulina:2018jjw}, including \nuc{Se}{76}, but suffered experimental shortcomings~\footnote{The authors have reported problems with the instrumentation used to record the timing information.}.
This paper presents a new determination of the muon lifetime in \nuc{Se}{76}, which is relevant for the ongoing \ac{onbb}-decay searches with \nuc{Ge}{76} of experiments like LEGEND~\cite{LEGEND:2025jwu}. 
The measurement principle and experimental setup are introduced in Sec.~\ref{sec:measurement}.
The analysis procedure and systematic uncertainties are discussed in Sec.~\ref{sec:analysis}. 
The results are presented in Sec.~\ref{sec:results} and the concluding discussion is provided in Sec.~\ref{sec:discussion}.
\section{\label{sec:measurement}Measurement}

This section outlines the measurement principle employed by the \ac{monument} collaboration and the experimental setup of the 2021 campaign, but is limited to the extent required to comprehend the results presented in this paper.
A detailed description of the experiment can be found in \cite{Araujo:2024mdu}.

\begin{figure}
\includegraphics[width=0.8\columnwidth]{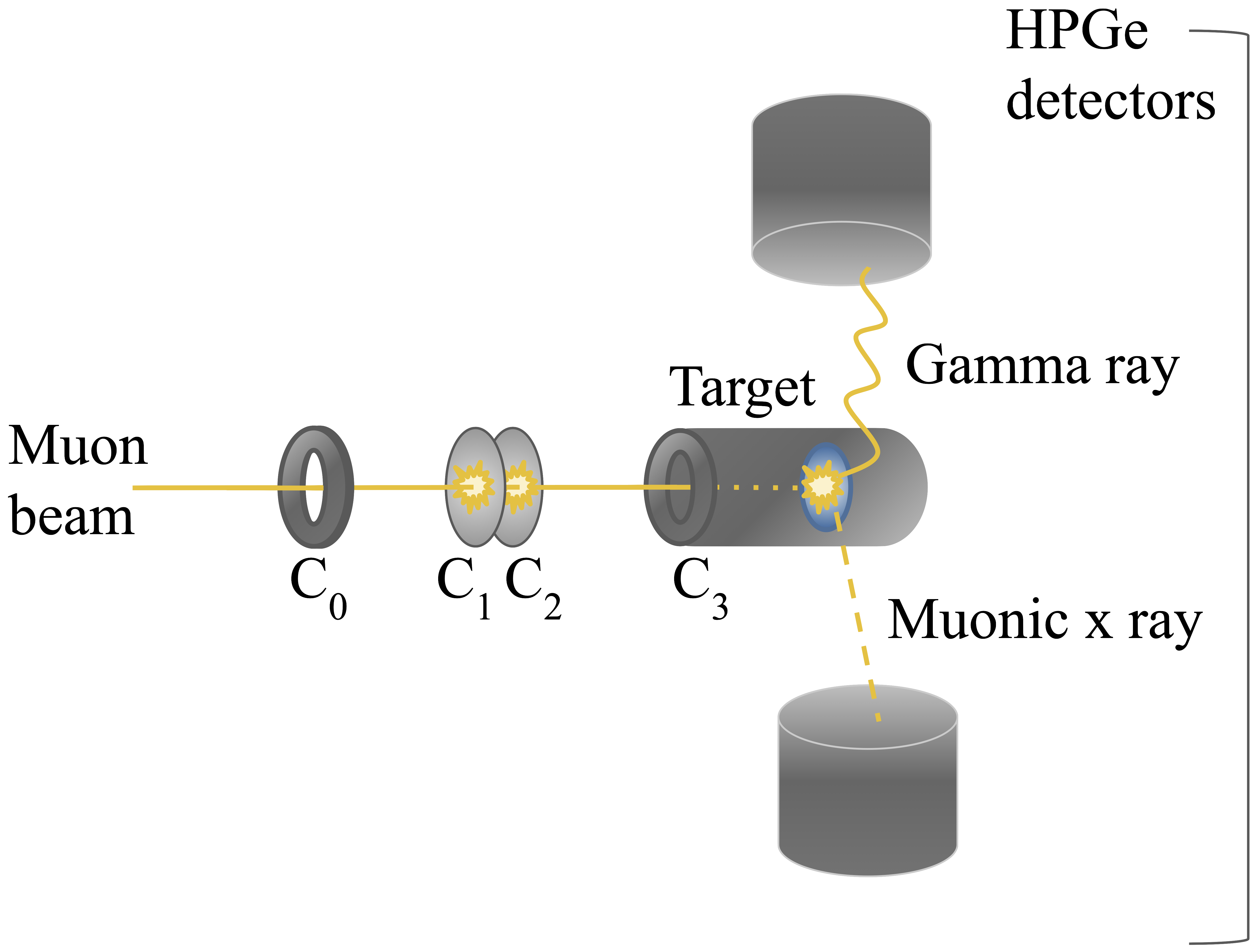}
\caption{\label{fig:measurement}   
Measurement principle of \acs{monument}.
No signal in the scintillator counter C\(_0\), coincident signals in C\(_1\) and C\(_2\), and no signal in C\(_3\) identify muons that stopped in the target, whereas the \acs{hpge} detector array provides spectroscopic information about the subsequent emission of $\upgamma$ and $\upmu$X rays
from the target.
}
\end{figure}

The \ac{monument} measurements were performed at the \(\pi\)E1 muon beam line of the \ac{psi}, providing negative muons with a momentum of about 38~MeV/c, which was selected to maximize the stopped muon rate.
A series of plastic scintillator counters, read out with \ac{pmt}, allowed to tag the path of the muons from the beam to the target.
It comprised four counters: a ring-shaped veto counter C\(_0\), two thin pass-through counters C\(_1\) and C\(_2\), and a cup-shaped veto counter C\(_3\).
Muons that stopped in the target produced no signal in C\(_0\), coincident signals in C\(_1\) and C\(_2\), and no signal in C\(_3\).
The typical rate of stopped muons was on the order of \(10^{4}\)~/s.
Spectroscopic information about the subsequent $\upgamma$ and muonic X ($\upmu$X)-ray emission was collected with an array of \ac{hpge} detectors surrounding the target chamber.
Fig.~\ref{fig:measurement} illustrates the measurement principle.

During the 2021 campaign, the \ac{hpge} detector array comprised eight detectors: two p-type \ac{bege} detectors with efficiencies of approximately 20~\%, and six p- or n-type \ac{coax} detectors with efficiencies ranging from 50~\% to 95~\%~\footnote{The quoted efficiencies correspond to the relative full-energy peak efficiency of the HPGe detector at 1332~keV, referenced to a 3\,$\times$\,3~inch NaI(Tl) detector under identical geometric conditions.} 
The total solid-angle coverage of the HPGe detector array was $<20$~\%. 
The \ac{bege} detectors delivered superior energy resolution, aiding the identification of spectral lines, while the \ac{coax} detectors offered better statistics and timing performance.

The data were collected with two independent \ac{daq} systems employing different triggering and recording schemes, one based on the \acs{midas} framework~\cite{midas:daq} and the other one based on an in-house development, called \acs{alpaca}~\cite{Schwarz:2024fei}.
In \acs{midas}, each \ac{hpge} detector and \ac{pmt} channel was configured to trigger independently, recording online energy and timing information based on a sampling frequency of 250~MHz. 
If a preceding trigger appeared within less than 1.4~µs, relevant \ac{pmt} information was not recorded due to the digitization time.
This increases the chance of missing true coincidences and recording random coincidences instead.

In \ac{alpaca}, the readout was initiated by the \ac{hpge} detectors, recording full waveforms of the triggering channel and all four \acp{pmt}.
For each channel, two waveform ranges were recorded.
For the \ac{hpge} detectors, a long \ac{lf} waveform from -10 to 10~\(\mu\)s at 15.625~MHz and a short \ac{hf} waveform from -1 to 1~\(\mu\)s at 125~MHz were acquired. 
For the \ac{pmt} channels, a long \ac{lf} waveform from -5.4 to 1.6~\(\mu\)s at 31.25~MHz and a short \ac{hf} waveform from -2 to 0~\(\mu\)s at 125~MHz were recorded.
Their ranges are defined relative to the online \ac{hpge} detector trigger signal.
This scheme allows offline reconstruction based on existing analysis tools~\cite{Agostini:2011xe}, but resulted in a shorter overall live time~\cite{Araujo:2024mdu}. 

Based on these two data streams, two independent analyses were carried out, further referred to as ALPACA and MIDAS analysis, allowing for a robust cross-validation of the new lifetime result, correcting the previous work presented in \cite{Zinatulina:2018jjw}.
The result presented in this paper is based on data collected between 29 October and 4 November 2021, with a total acquisition time of approximately 125~h in \ac{midas} and 126~h in \ac{alpaca}.
The two data sets are based on different data selection criteria, but partially share the same events.
The target was about 2~g of selenium powder enriched to 99.68~\% of \nuc{Se}{76}~\cite{Araujo:2024mdu}.
The second-most common isotope in the target was \nuc{Se}{77}, contributing 0.22~\%.
All other isotopes are present at the 0.01~\% level, or below.
The chemical purity was 99.96~\%.
\section{\label{sec:analysis}Analysis method}
The muon lifetime in \nuc{Se}{76} was determined from the time differences between the muons reaching the target and the subsequent $\upgamma$-ray emissions of the daughter nuclei after \ac{omc}.
This procedure relies on the assumption that the lifetimes of the excited nuclear states used in the analysis are negligible in comparison to the muon lifetime. 
The start time was provided by coincidences in the scintillator counters, while the stop time was provided by the \ac{hpge} detectors, identifying the $\upgamma$-ray emission with high resolution.
Timing effects were studied with $\upmu$X rays, which were emitted with negligible delays as the muons cascaded through the muonic orbits of the target atoms, yielding timing resolutions ranging between 20 to 100~ns~\cite{Araujo:2024mdu}.

The \ac{alpaca} and \ac{midas} data sets were analyzed using the same core methodology, but different implementations.  
A detailed description of the \ac{alpaca} analysis can be found in \cite{MondragonCortes:2025pim}.
In both cases, the starting point were two-dimensional histograms spanning the reconstructed \ac{hpge} detector energy \(E\) and the time \(t\) since the muon signal.
The start time was defined as the average time of the signals in the scintillator counters C\(_1\) and C\(_2\), considering only events with coincident hits in C\(_1\) and C\(_2\) and no signal in C\(_0\), within a 100~ns coincidence window.
The information provided by C\(_3\) was used to test for systematic uncertainties.

In \ac{alpaca}, the \ac{hpge} detector energy was obtained offline, applying optimized shaping parameters to the \ac{lf} waveforms, while the timing information was extracted from the \ac{hf} \ac{hpge} detector waveforms and the \ac{lf} \ac{pmt} waveforms~\cite{Araujo:2024mdu}.
This configuration was chosen, as it provides timing information that extends beyond the online \ac{hpge} detector trigger time.
The histograms were constructed with a binning of 0.2~keV and 32~ns, only considering events with single \ac{pmt} signals in a \(\pm0.8\)~µs window, covering $>99$~\% of all true coincidences.
Using this procedure, random coincidences are negligible~\cite{Araujo:2024mdu}. 
By construction, only events where the true coincidence was missed and a false positive muon signal was recorded could enter the histograms accidentally.
Random coincidences alone would not suffice the single-signal criterion, whenever the true coincident signal is recorded. 

\begin{figure}
\includegraphics[width=1\columnwidth]{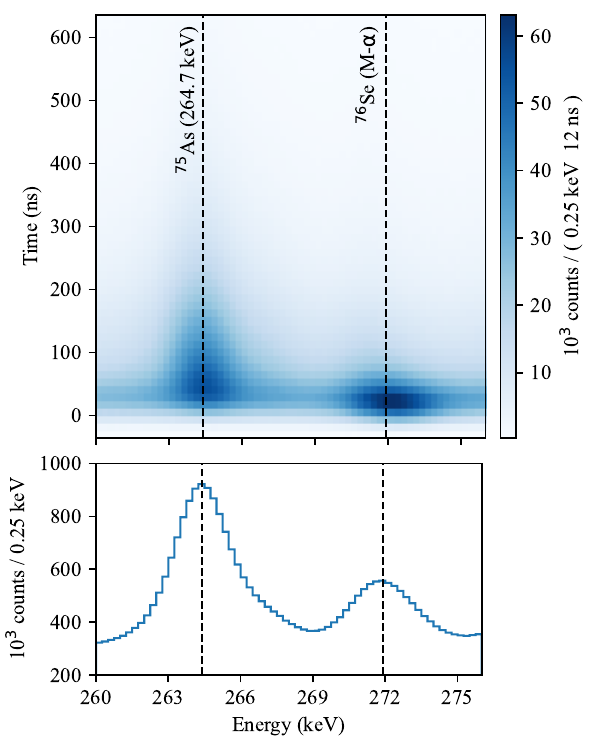}
\caption{\label{fig:histogram} 
Fragment of a two-dimensional histogram based on the \ac{midas} data, spanning the \ac{hpge} detector energy and the time since the muon signal.
The muon lifetime becomes apparent in the exponential extent of the $\upgamma$-ray lines towards later times, while the $\upmu$X rays appear prompt, evidencing the timing resolution.
The data was recorded with detector 5.
}  
\end{figure}

In MIDAS, the histograms are based on the online energy and timing information, using a binning of 0.25~keV in energy and 12~ns in time, spanning -2 to 10~µs.
To counteract possible distortions due to the individual dead time of each \ac{pmt} channel, all muon signals falling into this window were included, generating a flat contribution of random coincidences. 
Fig.~\ref{fig:histogram} shows a fragment of a two-dimensional histogram.

\subsection{\label{sec:time-profile-extraction}Time profile extraction}

The muon lifetime was determined individually for the different \ac{hpge} detectors and for different gamma lines emitted after \ac{omc}, using two analysis steps: the time profile extraction and the lifetime fit.
The time profiles represent the evolution of the \ac{fep} intensities over the time slices of the two-dimensional histograms. 

\begin{figure*}
\includegraphics[width=1\textwidth]{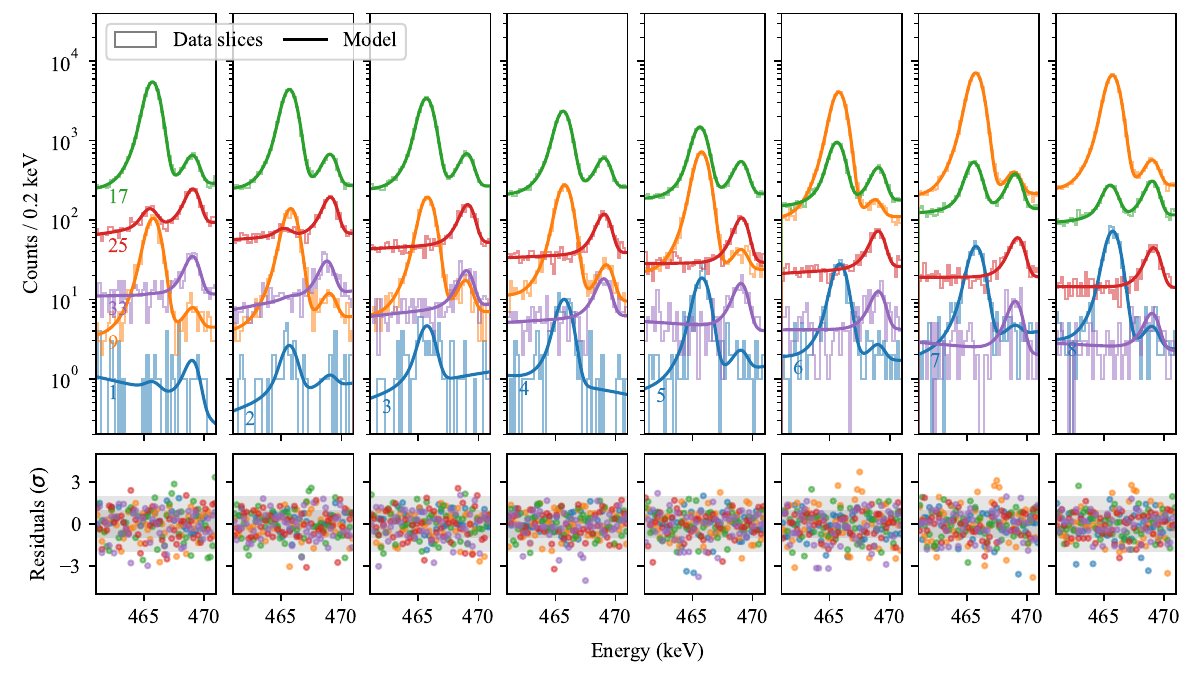}
\caption{\label{fig:time-profile-extraction} 
Time profile extraction in \ac{alpaca}. 
The intensities were extracted by a combined fit to all 40 time slices, each of which is 32~ns long, using shared \ac{fep} parameters, but individual backgrounds.
The slices are shown in consecutive order, using 8 windows and 5 colors.
Their indices indicate the slice number.
The $\upmu$X ray at 466~keV dominates the earlier times slices, whereas the $\upgamma$-ray line at 469~keV prevails until later times.
The data was recorded with detector 6, which achieves proper separation of the two lines.
}  
\end{figure*}

In \ac{alpaca}, the \acp{fep} were modeled with a standard two component function \(f_{\mathrm{FEP}}(E, \ldots)\), consisting of a Gaussian bulk and an exponentially modified Gaussian low-energy tail, attributed to incomplete charge collection and soft pile-up~\cite{GERDA:2021pcs}:
\begin{equation}
\begin{split}
f_{\mathrm{FEP}}(E; \ldots) = \frac{1 - \alpha}{\sqrt{2\pi}\sigma} 
\exp\left( -\frac{(E-\mu)^2}{2\sigma^2} \right) + \\
\quad \frac{\alpha}{2\beta} 
\exp\left( \frac{E-\mu}{\beta} + \frac{\sigma^2}{2\beta^2} \right) 
\operatorname{erfc}\left( \frac{E-\mu}{\sqrt{2}\sigma} + \frac{\sigma}{\sqrt{2}\beta} \right)
\end{split}
\label{eq:fep}
\end{equation}
where \(\mu\) is the peak position, \(\sigma\) is the width, \(\alpha\) regulates the ratio between bulk and tail, and \(\beta\) defines the exponential deficit of the tail.
The background was modelled as a first order polynomial, which was found sufficient to describe the data locally within the fit windows, typically limited to \(\pm5\)~keV around the \acp{fep}:
\begin{equation}
    f_{\mathrm{bkg}}(E; \ldots) = p_0 + p_1 \cdot E
\end{equation}
In cases with more than one line within a fit window, the shape parameters were shared among all \acp{fep}:
\begin{eqnarray}
  &f(E; n_{ij}, \ldots) = \sum_i n_{ij} \cdot f_{\mathrm{FEP}}(E; \mu_i, \ldots)+\nonumber\\
  &f_{\mathrm{bkg}}(E; \ldots)
  \label{eq:fep}
\end{eqnarray}
Based on this model, the intensities \(n_{ij}\) of all \acp{fep} \(i\) were extracted by a combined maximum likelihood fit of all time slices \(j\), each of which was 32 ns long, covering a time window of \(\pm0.64\)~µs, considering Poisson statistics and using shared shape parameters, but individual backgrounds.
The time profiles were then propagated to the lifetime fits, together with the marginalized covariance matrix encoding the correlations due to the common shape parameters.
The correlations coefficients between the intensities extracted from different time slices reached values of up to several 10~\%.
Fig.~\ref{fig:time-profile-extraction} shows a result of this analysis step.

In \ac{midas}, the time profile extractions were carried out in two steps.
In a first step, the total energy projections were fit to obtain the \ac{fep} parameters, while then in a second step, each time slice was fit individually to extract the \ac{fep} intensity evolution, keeping the \ac{fep} parameters fixed.
The fits were carried out as binned \(\chi^2\) fits, justified by the overall larger statistics and random coincidences, dominating the later time slices.
The model configuration was essentially the same, except for an additional high-energy tail added to Eq.~\ref{eq:fep}, which was attributed to \ac{hpge} detector signals sitting on recovering baselines after test pulse injections, that were removed in the \ac{alpaca} data stream by quality cuts~\cite{Araujo:2024mdu}.

\subsection{\label{sec:lifetime-fit}Lifetime fit}
The muon lifetime was then extracted from the time profiles of the $\upgamma$-ray lines, consisting of a prompt timing response \(f_{\mathrm{prompt}}(t; \ldots)\) convolved with an exponential decay, encoding the muon lifetime \(\tau\):
\begin{equation}
  f(t; \tau, \ldots) = f_{\text{prompt}}(t; \ldots) \circledast \exp\left(-\frac{t}{\tau}\right).
  \label{eq:time-profile2}
\end{equation}
The prompt response describes the underlying time distribution of prompt events, containing all timing effects of the system, such as the time resolutions, delays between the different channels and reconstruction artifacts, such as trigger walks.
As the $\upmu$X rays were emitted with negligible delay, their time profiles served as reliable proxies of the prompt timing response.
The time resolutions range between 20~ns and 100~ns, depending on the \ac{hpge} detector and $\upgamma$-ray line energy~\cite{Araujo:2024mdu}.

\begin{figure}
\includegraphics[width=1\columnwidth]{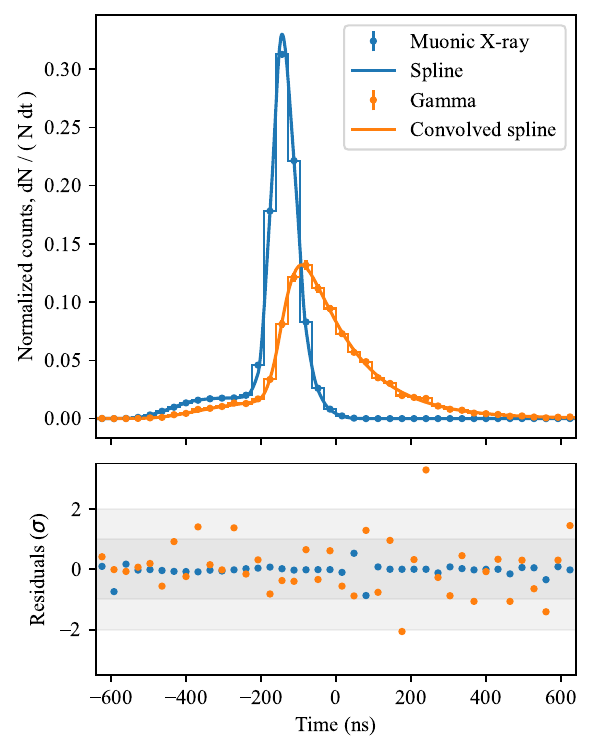}
\caption{\label{fig:lifetime-fit}
Lifetime fit in \ac{alpaca}.
The lifetime was obtained by a combined fit of the $\upmu$X ray, providing the prompt timing response, and the $\upgamma$-ray line, whose exponential decay encodes the muon lifetime.
The data correspond to the $\upmu$X ray line at 456~keV and $\upgamma$-ray line at 419~keV recorded with detector 4.
} 
\end{figure}

In \ac{alpaca}, the muon lifetime was extracted by combined fits of the time profiles of neighboring $\upgamma$ and $\upmu$X-ray lines, featuring similar timing effects. 
As no analytical expression was found to consistently describe the $\upmu$X-ray time profiles across all detectors, an interpolating monotonic cubic spline through virtual intensity points placed at the bin centers was used~\cite{PCHIP}, providing a continuous description of the $\upgamma$ and $\upmu$X-ray lines, with and without the convolution with the exponential decay.
The fits were performed as combined shape-only binned \(\chi^2\) fits of the full time profiles, using the intensities and covariance matrices obtained in the time profile extraction.
Fig.~\ref{fig:lifetime-fit} shows a result of this analysis step.

\begin{figure}
\includegraphics[width=1\columnwidth]{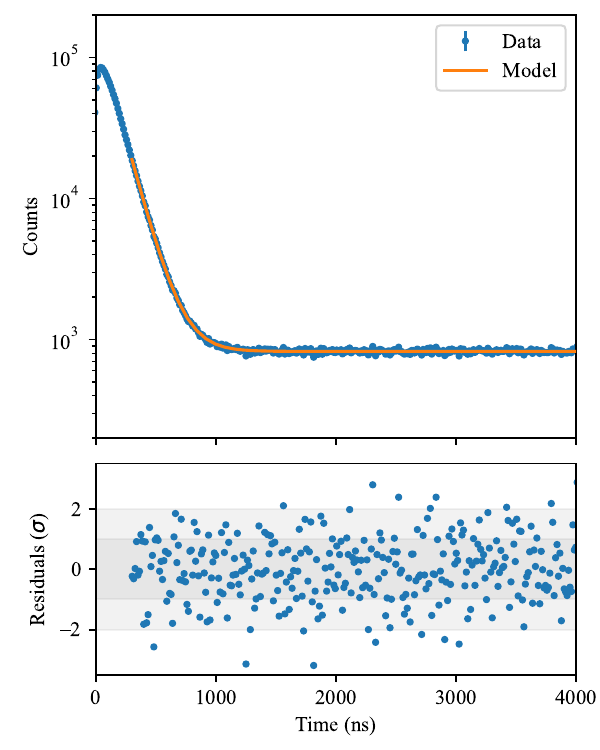}
\caption{\label{fig:lifetime-fit2}
Lifetime fit in \ac{midas}.
The lifetime was obtained by a fit to the tail of the $\upgamma$-ray line intensities using an exponential decay on a constant random coincidence background.
The data corresponds to the $\upgamma$-ray line at 264~keV recorded with detector 7.
} 
\end{figure}

In \ac{midas} the lifetime fits were performed differently, starting the fit from a certain time point, when the prompt response has settled, where the time profile of $\upgamma$-ray lines can be described by an exponential decay, sitting on a constant background due to random coincidences.
Accordingly, no nearby $\upmu$X rays are required, and the muon lifetime was extracted by fits of the $\upgamma$-ray time profile tails with ranges starting as early as 200~ns, and as late as 500~ns, carried out as \(\chi^2\) fits, using the individual intensities and uncertainties obtained in the time profile extraction.
This approach relies on a sufficiently late starting point of the fit, which is considered a systematic uncertainty.
Fig.~\ref{fig:lifetime-fit2} shows a result of this analysis step.

\subsection{\label{sec:systematics}Systematic uncertainties}

Most generally, any unaccounted distortion of the time profiles or shortcomings in the lifetime fits can lead to a systematic bias on the muon lifetime.
The systematic effects considered for the results presented in this paper are related to features in the \ac{hpge} detector charge collection, the background model in the time profile extraction, the energy dependence of the prompt timing response, and the lifetime fit methods themselves, mostly concerning the choice of binnings and fit ranges.

The time profile extraction relies on the assumption that the \ac{fep} parameters are the same for every time slice.
This assumption is broken for events featuring an energy deficit due to charge trapping, which is more likely for events with larger drift times of the charge carriers in the \acs{hpge} detector, that were reconstructed at later times by the triggering algorithms.
This effect became especially apparent for $\upmu$X rays, where the centroid of correlated \ac{fep} events shifts to lower energies at later times.
In \ac{alpaca}, this effect was taken into account in the time profile extractions.
The \ac{fep} positions and tail fractions in each time slice \(j\) were given additional flexibility by individual parameters \(\Delta\mu_j\) and \(\Delta\alpha_j\), bound to the central parameters \(\mu\) and \(\alpha\) by Gaussian pull terms.
The width of the pull terms was set to weak values of 1~keV and 5~\% for time slices where the statistics was considered sufficient to constrain the actual \ac{fep} parameters, whereas strong pull terms of 0.1~keV and 1~\% were adapted for low statistics slices, but adding an additional uncertainty on the resulting \ac{fep} intensities in quadrature, corresponding to the statistical uncertainty of the underlying background.
This procedure of including additional quasi-free nuisance parameters increased the muon lifetime uncertainties by about 10~\%, but was shown to recover the true value in \ac{mc} simulation studies, where the energy deficit leading to the low-energy tail is correlated with later times and was hence considered uncorrelated across the individual measurements~\cite{MondragonCortes:2025pim}.
In \ac{midas}, this effect was taken into account as a correlated uncertainty of 0.5~ns, common to all results, estimated from a similar set of \ac{mc} simulations, where no additional flexibility is given to the fits.

Another assumption in the time profile extraction is that the background in each slice can be described by a first-order polynomial, meaning a linear function.
This simplification can generate a bias.
If the true background has a convex shape, actual background counts could have been attributed as \ac{fep} intensities, leading to overestimations depending on the background level.
The opposite applies to a true background of concave shape.
As the background does not necessarily follow the same time dependence as the signal, missmodelling of the background can distort the time profile.
This effect was included as a correlated uncertainty of 0.2~ns, common to all \ac{alpaca} and \ac{midas} results, estimated from the uncertainty that the next order polynomial could contribute.

\begin{table*}
\caption{\label{tab:systematics}Systematic uncertainties. 
Apart from the background shape uncertainty, the systematic uncertainties and their treatment differ in the two analyses.
Changes in the line positions, stemming from \acs{hpge} detector charge collection effects, as well as the energy dependence of the prompt timing response are taken into account by additional nuisance parameters in the \acs*{alpaca} fits.
The method uncertainties are considered uncorrelated between the two analyses.}
\begin{ruledtabular}
\begin{tabular}{l c c}
\textbf{Effect} & \textbf{ALPACA} & \textbf{MIDAS} \\
\hline
Line positions, \acs*{hpge} detector charge collection & additional parameters & 0.5~ns \\
Background shape & \multicolumn{2}{c}{0.2~ns} \\
Prompt timing response, energy dependence  & additional parameters & not applicable \\
Method & 0.6~ns & 0.4~ns \\
\end{tabular}
\end{ruledtabular}
\end{table*}

The exact prompt timing response depends on the energy, as higher-energy events are typically reconstructed with better time resolution.
This breaks the assumption that nearby $\upgamma$ and $\upmu$X-ray lines can be described by the same prompt response taken in the combined lifetime fit in \ac{alpaca}.
To account for this effect, the description of the timing response was given additional freedom before its application to the $\upgamma$-ray line, using a simple transformation acting on the virtual intensity points of the spline, implemented as a linear stretch starting at an anchor point.
It was shown that this method can translate the timing response from one $\upmu$X-ray line to another~\cite{MondragonCortes:2025pim}.
The additional nuisance parameters of the transformation, bound by weak pull terms, increased the muon lifetime uncertainties by about a factor of 2, but were assumed uncorrelated across all measurements, as even fully free parameters did not change the outcome numerically.

The general ability to reconstruct the true muon lifetime with the \ac{alpaca} and \ac{midas} lifetime fit methods was studied in \ac{mc} simulations and by performing fits under varying configurations, such as binning, coincidence criteria, and \ac{hpge} detector time reconstruction algorithms.
The overall method uncertainty is 0.6~ns in \ac{alpaca}, attributed mostly to the binning of 32~ns, and 0.4~ns in \ac{midas}, mostly attributed to the choice of the start time of the tail fit.
They were considered correlated among all measurements obtained with the same analysis implementation, but uncorrelated across the two methods, even though simulations suggested a slight anti-correlation from which the overall uncertainty could profit.
Tab.~\ref{tab:systematics} summarizes the systematic uncertainty budget.
\section{\label{sec:results}Results}

In total, 70 statistically independent values from different gamma lines and \ac{hpge} detectors were obtained, 25 in \ac{alpaca} and 45 in \ac{midas}.
They originated from transitions in \nuc{As}{76}, the direct product of \ac{omc} on \nuc{Se}{76}, and \nuc{As}{74} and \nuc{As}{75}, which appear after neutron emission. 
Tab.~\ref{tab:taus} in Appx.~\ref{sec:appendix} summarizes these results.

\begin{figure}
\includegraphics[width=1\columnwidth]{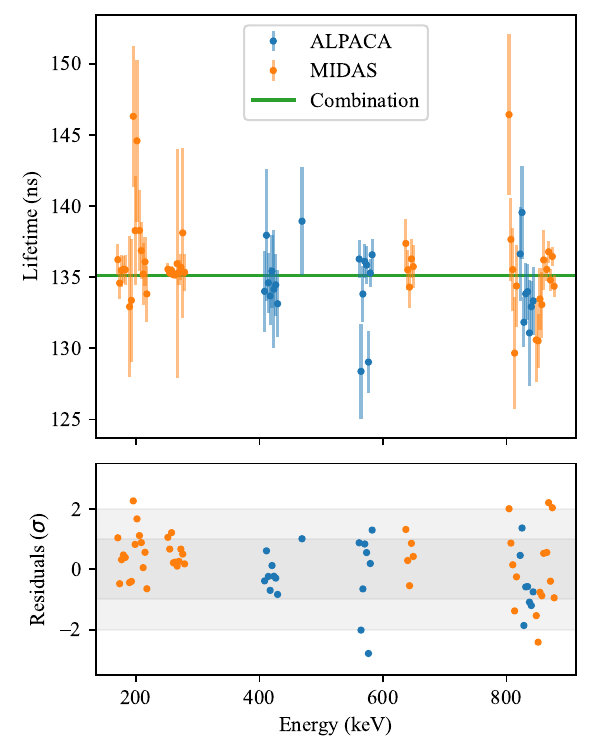}
\caption{\label{fig:lifetimes} 
Individual muon lifetime results and combination. 
In total 70 individual measurements were carried out, 25 based on \ac{alpaca} data and 45 with \ac{midas} data.
They are plotted at the corresponding gamma line energy, but shifted among the different \ac{hpge} detectors to aid visibility.
Only uncorrelated uncertainties are plotted, but all correlated uncertainties are considered in the combination.}  
\end{figure}

In \ac{alpaca} the lines were chosen due to their proximity to muonic x-ray lines, providing the prompt timing profile in the combined lifetime fit.
They correspond to 4 (gamma, muonic x-ray) combinations at (419, 456), (469, 466), (572, 614) and (822, 752)~keV, all originating from \nuc{As}{75} and the muonic x-ray L series. 
The analysis was performed for all 8 \ac{hpge} detectors, except for the (469, 466)~keV combination, which was only analyzed for a \ac{bege} detector with sufficient energy resolution to properly separate the two lines.

In order to avoid data sets with overlapping statistics, the \ac{midas} analysis provided results for another 9 gamma lines at 183, 202, 206, 264, 267, 643, 816, 860, and 865~keV and 5 \ac{hpge} detectors, excluding 3 \ac{hpge} detectors with insufficient energy or time resolutions.

The combination of all values, taking into account systematic uncertainties as covariances, results in a muon lifetime in \nuc{Se}{76} of
\begin{equation*}
    \tau = (135.1\pm0.5) \, \text{ns}.
\end{equation*}
The \(\chi^2\) of the combination is 70.1 for 69 degrees of freedom, which corresponds to a p-value of 0.4.
The systematic uncertainties dominate. 
The pure statistical uncertainty lies slightly below 0.1~ns.
The individual results are \((134.8\pm0.7)\)~ns for \ac{alpaca} and \((135.4\pm0.7)\)~ns for \ac{midas}.
Their agreement lends further credibility to this new result.
The tail fit analysis applied to the \ac{alpaca} data resulted in \((133.9\pm0.8)\)~ns, which also agrees reasonably well with the result.
Similarly, the \ac{midas} analysis chain, applied to the \ac{alpaca} data, produced consistent results.
Fig.~\ref{fig:lifetimes} shows the individual values and the combined result.

\section{\label{sec:discussion}Discussion}

In summary, the improved accuracy through the comprehensive treatment of systematic uncertainties and the cross-check provided by the two independent data streams and analyses, lends credibility to this first \acs{monument} physics result, correcting the muon lifetime in \nuc{Se}{76} from \((148.5\pm0.1)\)~ns presented in \cite{Zinatulina:2018jjw} to the new value of \((135.1\pm0.5)\)~ns.
A first comparison with theoretical calculations of the total capture rate, \(\Lambda_{\textrm{cap}}\), revealed interesting features.
The inverse of the lifetime is the sum of the total rate of muon captures and decays:
\begin{equation}
     \frac{1}{\tau} = \Lambda_{\textrm{cap}} + \mathcal{H}\Lambda_{\textrm{dec}},
    \label{eq:total-rate}
\end{equation}
where $\Lambda_{\textrm{dec}}$ is the free muon decay rate of \(0.455\times10^{6}\)~/s~\cite{Mukhopadhyay1980}, correct by a Huff factor, \(\mathcal{H}\), of 0.955~\cite{Huff1961,Suzuki:1987jf}.
Tab.~\ref{tab:lifetimes} summarizes all available lifetime values for \nuc{Se}{76}.

\begin{table}
\caption{\label{tab:lifetimes}
Muon lifetimes in \nuc{Se}{76}.
The new experimental result agrees with the \acs{qrpa} calculation of \cite{Simkovic:2020geo} for unquenched axial-vector coupling \(g_A\), as well as with the semi-empirical calculations of Primakoff or Goulard-Primakoff type.
The theoretical lifetimes were derived from the capture rates using Eq.\ref{eq:total-rate}.
The parameters used in the Primakoff and Goulard–Primakoff calculations were taken from \cite{Measday:2001yr}, resulting in a range of lifetimes for the latter.} 
\begin{ruledtabular}
\begin{tabular}{l c c}
Source & Lifetime (ns) & \(g_A\) \\
\hline
\multicolumn{3}{l}{\textbf{Phenomenological models}}  \\
\hline
pn\acs*{qrpa}~\cite{Jokiniemi:2019nne} & 59.4  & 0.80 \\
\acs*{qrpa}~\cite{Simkovic:2020geo}    & 254.1 & 0.80 \\
                                       & 192.1 & 1.00 \\  
                                       & 134.5 & 1.27 \\
\hline
\multicolumn{3}{l}{\textbf{Semi-empirical models}} \\
\hline
Primakoff~\cite{Primakoff:1959fs}       & 135.2 & \\
Goulard-Primakoff~\cite{Goulard:1974zz} & 115.1--135.3 & \\
Fujii-Primakoff~\cite{Simkovic:2020geo} & 196.3 & \\
\hline
\multicolumn{3}{l}{\textbf{Experiments}} \\
\hline
This work                            & $135.1\pm0.5$ \\
Other work~\cite{Zinatulina:2018jjw} & $148.5\pm0.1$ \\
\end{tabular}
\end{ruledtabular}
\end{table}

To this day, two phenomenological calculations of the total muon capture rate in \nuc{Se}{76} are available.
Both are based on the \ac{qrpa}.
The proton-neutron \acs{qrpa} (pn\acs{qrpa}) calculations presented in \cite{Jokiniemi:2019nne} provide a much higher capture rate, translating into a much shorter muon lifetime than the experimental result, even though they incorporate a strongly quenched axial-vector coupling constant of \(g_A = 0.8\).
Conversely, the \acs{qrpa} calculations presented in \cite{Simkovic:2020geo}, which were carried out as a response to \cite{Jokiniemi:2019nne} with different quenching factors, reproduce the experimental result when the unquenched value of \(g_A = g_A^{\mathrm{free}} = 1.27\) is used.
This indicates that little to no quenching is required within the \ac{qrpa} framework to describe the \acs{omc} total capture rate, which is consistent with earlier works~\cite{Zinner:2006jv, Marketin:2008ei}.
Further calculations, including the \ac{onbb} decay \acp{nme} using the very same configurations, also using other phenomenological or ab-initio models, would be of great interest to the community.

The experimental result also agrees with semi-empirical calculations, especially when using the Primakoff~\cite{Primakoff:1959fs} or Goulard-Primakoff~\cite{Goulard:1974zz} formalism, using an effective nuclear charge of Z$_{\text{eff}}=23.405$ for \nuc{Se}{76}~\cite{Simkovic:2020geo}.
Further details on these comparisons can be found in \cite{MondragonCortes:2025pim}.
This suggests that the mere comparison of phenomenological calculations to semi-empirical predictions could be an interesting avenue to benchmark nuclear physics models.

The muon lifetime in \nuc{Se}{76} is the first physics result obtained by the \acs{monument} collaboration, building up on an exploratory measurement with a modest array of \ac{hpge} detectors.
The partial capture rates, meaning the individual capture strengths to excited \nuc{As}{76} states, remain subject to further analysis efforts and could benefit from experimental improvements.
This will require careful reconstruction of de-excitation cascades and the consideration of competing processes~\cite{Grabmayr:2025mfd}, especially given the low detection efficiency for high-energy $\upgamma$ rays. 
Future measurements could profit from Compton suppression shields, aiding the detection of weak transitions, increased detection efficiency and sophisticated coincidence analyses.

\begin{acknowledgments}
This work was supported by the RFBR and DFG under project number 21-52-12040 and DFG Grant 448829699, by the U.S. DOE under Grant No. DE-SC0019261, by FWO-Vlaanderen (Belgium), and by BOF KU Leuven under contract No. C14/22/104.  
We thank the Paul Scherrer Institute for enabling these measurements at the High-Intensity Proton Accelerator of the Swiss Infrastructure for Particle Physics.  
We thank C.~Petitjean and F.~Wauters for their support during the 2021 measurement campaign. 
E.\,Mondrag\'on thanks S.~Mertens and the Max-Planck-Institut für Kernphysik in Heidelberg for their support during the preparation of this manuscript.
\end{acknowledgments}

\bibliography{references}


\onecolumngrid
\appendix
\clearpage
\section{\label{sec:appendix}Individual lifetime results}

\vspace{-1\baselineskip}
\begin{table*}[h!]
\centering
\caption{\label{tab:taus}Individual lifetime results for different $\upgamma$-ray lines and \acs*{hpge} detectors. 
The gamma lines at energies \(E\) of 183, 202, 206, and 267~keV originate from \nuc{As}{74}; those at 264, 419, 469, 572, 816, 822, 860, and 865~keV from \nuc{As}{75}; and that at 643~keV from \nuc{As}{76}. 
The superscript denotes \ac{alpaca} (A) or \ac{midas} (M) analysis. 
The detector numbers \# of 1, 3, 4, and 5 correspond to n-type \ac{coax} detectors; 2 and 6 are \ac{bege} detectors; and 7--8 are p-type \ac{coax} detectors.}
\begin{ruledtabular}
\begin{tabular}{ccc@{\hspace{0.3cm}}|@{\hspace{0.3cm}}ccc@{\hspace{0.3cm}}|@{\hspace{0.3cm}}ccc}
 \textbf{\(E\)~(keV)} & \# & \textbf{Lifetime~(ns)} &
 \textbf{\(E\)~(keV)} & \# & \textbf{Lifetime~(ns)} &
 \textbf{\(E\)~(keV)} & \# & \textbf{Lifetime~(ns)}\\
\hline
183$^\text{M}$  & 1    & 136.2 $\pm$ 1.1  & 419$^\text{A}$ & 1 & 134.0 $\pm$ 2.9 & 816$^\text{M}$ & 5    & 136.5 $\pm$ 0.7 \\
                & 3    & 134.6 $\pm$ 1.1  &                & 2 & 137.9 $\pm$ 4.7 &                & 7    & 134.4 $\pm$ 0.8 \\
                & 4    & 135.5 $\pm$ 1.1  &                & 3 & 134.6 $\pm$ 2.1 & 822$^\text{A}$ & 1    & 136.6 $\pm$ 3.3 \\
                & 5    & 135.6 $\pm$ 0.9  &                & 4 & 133.7 $\pm$ 2.0 &                & 2    & 139.5 $\pm$ 3.3 \\
                & 7    & 135.5 $\pm$ 1.1  &                & 5 & 135.4 $\pm$ 2.6 &                & 3    & 131.8 $\pm$ 1.8 \\
202$^\text{M}$  & 1    & 132.9 $\pm$ 5.0  &                & 6 & 134.2 $\pm$ 4.1 &                & 4    & 133.8 $\pm$ 2.2 \\
                & 3    & 133.4 $\pm$ 4.3  &                & 7 & 134.5 $\pm$ 2.2 &                & 5    & 134.0 $\pm$ 2.0 \\
                & 4    & 146.3 $\pm$ 5.0  &                & 8 & 133.1 $\pm$ 2.4 &                & 6    & 131.1 $\pm$ 3.7 \\
                & 5    & 138.3 $\pm$ 3.9  & 469$^\text{A}$ & 6 & 138.9 $\pm$ 3.8 &                & 7    & 132.9 $\pm$ 1.8 \\
                & 7    & 144.6 $\pm$ 5.7  & 572$^\text{A}$ & 1 & 136.3 $\pm$ 1.3 &                & 8    & 133.3 $\pm$ 2.4 \\
206$^\text{M}$  & 1    & 138.3 $\pm$ 2.9  &                & 2 & 128.4 $\pm$ 3.3 & 860$^\text{M}$ & 1    & 130.6 $\pm$ 2.9 \\
                & 3    & 136.9 $\pm$ 2.0  &                & 3 & 133.8 $\pm$ 2.0 &                & 3    & 130.5 $\pm$ 1.9 \\
                & 4    & 135.2 $\pm$ 2.2  &                & 4 & 136.1 $\pm$ 1.2 &                & 4    & 133.5 $\pm$ 2.2 \\
                & 5    & 136.1 $\pm$ 1.7  &                & 5 & 135.9 $\pm$ 1.4 &                & 5    & 133.1 $\pm$ 2.3 \\
                & 7    & 133.8 $\pm$ 2.0  &                & 6 & 129.0 $\pm$ 2.2 &                & 7    & 136.2 $\pm$ 2.1 \\
264$^\text{M}$  & 1    & 135.6 $\pm$ 0.4  &                & 7 & 135.3 $\pm$ 1.0 & 865$^\text{M}$ & 1    & 135.5 $\pm$ 0.8 \\
                & 3    & 135.3 $\pm$ 0.3  &                & 8 & 136.6 $\pm$ 1.1 &                & 3    & 136.8 $\pm$ 0.8 \\
                & 4    & 135.5 $\pm$ 0.3  & 643$^\text{M}$ & 1 & 137.4 $\pm$ 1.7 &                & 4    & 134.8 $\pm$ 0.8 \\
                & 5    & 135.2 $\pm$ 0.3  &                & 3 & 135.5 $\pm$ 1.4 &                & 5    & 136.5 $\pm$ 0.7 \\
                & 7    & 135.2 $\pm$ 0.3  &                & 4 & 134.3 $\pm$ 1.5 &                & 7    & 134.4 $\pm$ 0.8 \\
267$^\text{M}$  & 1    & 135.9 $\pm$ 8.0  &                & 5 & 136.3 $\pm$ 1.4 & \\
                & 3    & 135.3 $\pm$ 0.9  &                & 7 & 135.7 $\pm$ 1.5 & \\
                & 4    & 135.7 $\pm$ 0.9  & 816$^\text{M}$ & 1 & 146.4 $\pm$ 5.7 & \\
                & 5    & 138.1 $\pm$ 6.0  &                & 3 & 137.7 $\pm$ 2.9 & \\
                & 7    & 135.3 $\pm$ 1.3  &                & 4 & 135.5 $\pm$ 2.9 & \\
    \end{tabular}
\end{ruledtabular}
\end{table*}
\twocolumngrid

\end{document}